\documentclass[letterpaper, 10 pt, conference]{ieeeconf}
\IEEEoverridecommandlockouts                             
\usepackage{subfiles}
\usepackage{graphics} 
\usepackage{epsfig} 
\usepackage{mathptmx} 
\usepackage{times} 

\usepackage{amsmath} 
\usepackage{amssymb}  
\usepackage{bbm}
\usepackage{xcolor}
\usepackage{graphicx}  
\usepackage{amsthm}
\usepackage{setspace}
\usepackage{soul}
\usepackage[ruled,vlined]{algorithm2e}
\usepackage{gensymb}
\usepackage{url}
\usepackage{multirow}

\title{\LARGE \bf
State Estimation for Legged Robots \\ Using Contact-Centric Leg Odometry
}

\author{Shuo Yang$^{1}$, Hans Kumar$^{1}$, Zhaoyuan Gu$^{1}$, Xiangyuan Zhang$^{1,2}$, Matthew Travers$^{1}$ and Howie Choset$^{1}$
\thanks{$^{1}$Biorobotics Laboratory, The Robotics Institute, Carnegie Mellon University, Pittsburgh, PA 15213, USA
{\tt\small \{shuoyang, hansk, zhaoyuan, xiangyua, mtravers, choset\}@andrew.cmu.edu}}
\thanks{$^{2}$Coordinated Science Laboratory, University of Illinois at Urbana-Champaign, Urbana, IL 61801, USA}%
}

\begin{document}

\maketitle
\thispagestyle{empty}
\pagestyle{empty}

\begin{abstract}

Our goal is to send legged robots into challenging, unstructured terrains that wheeled systems cannot traverse. Moreover, precise estimation of the robot's position and orientation in rough terrain is especially difficult. To address this problem, we introduce a new state estimation algorithm which we term Contact-Centric Leg Odometry (COCLO). This new estimator uses a Square Root Unscented Kalman Filter (SR-UKF) to fuse multiple proprioceptive sensors available on a legged robot. In contrast to IMU-centric filtering approaches, COCLO formulates prediction and measurement models according to the contact status of legs. Additionally, COCLO has an indirect measurement model using joint velocities to estimate the robot body velocity. In rough terrain, when IMUs suffer from large amounts of noise, COCLO's contact-centric approach outperforms previous IMU-centric methods. To demonstrate improved state estimation accuracy, we compare COCLO with Visual Inertial Navigation System (VINS), a state-of-the-art visual inertial odometry \cite{qin2018vins}, in three different environments: flat ground, ramps, and stairs. COCLO achieves better estimation precision than VINS in all three environments and is robust to unstable motion. Finally, we also show that COCLO and a modified VINS can work in tandem to improve each other's performance.

\end{abstract}

\section{Introduction}
We introduce a new state estimator for legged robots with low grade proprioceptive sensors operating in rough terrain. Designing such a state estimator is challenging because legged systems have many degrees of freedom and their interaction with the environment can be difficult to model. 
One major challenge in designing estimation algorithms for legged robots is addressing high energy noise upon foot contact with the ground. These contact impacts induce significant noise to the accelerometer readings that is difficult to eliminate, as shown in Figure \ref{fig:acc_noise}, making traditional IMU-centric estimation methods less accurate. Therefore, we need to introduce a different state estimation approach to tackle this problem. 

In this paper, we propose a legged robot state estimation algorithm termed Contact-Centric Leg Odometry (COCLO). COCLO fuses information from a low grade IMU, leg joint encoders, and torque sensors into an SR-UKF to estimate the full state of a legged robot where the state variables include center of mass (CoM) position and velocity, orientation quaternion, gyroscope bias, accelerometer bias, foot positions, and contact status of each foot. 

\begin{figure} 
    \centering
    \vspace{0.3cm}
       \includegraphics[width = 4.26cm, height = 3.5cm]{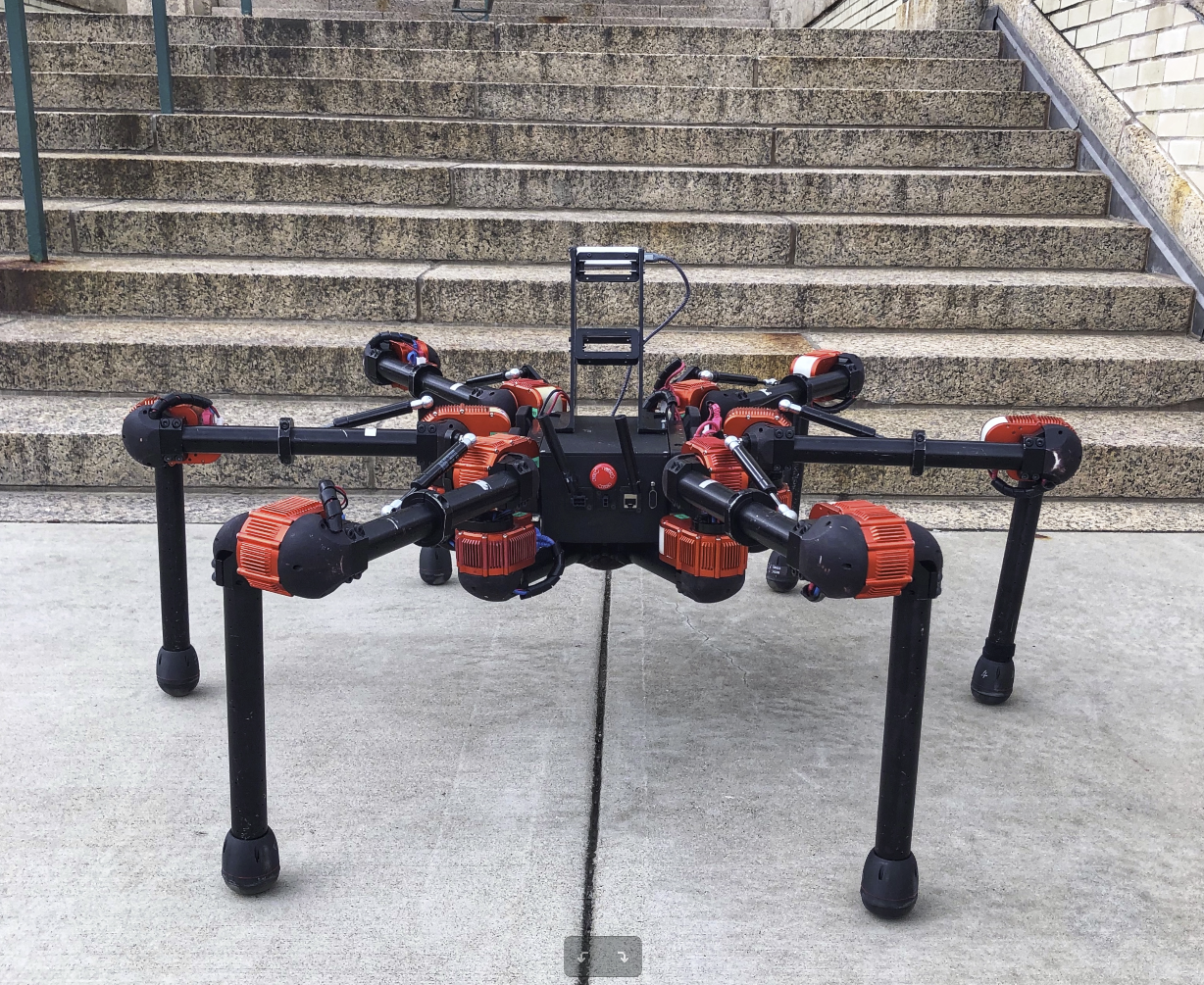}
       \includegraphics[width = 4.26cm, height = 3.5cm]{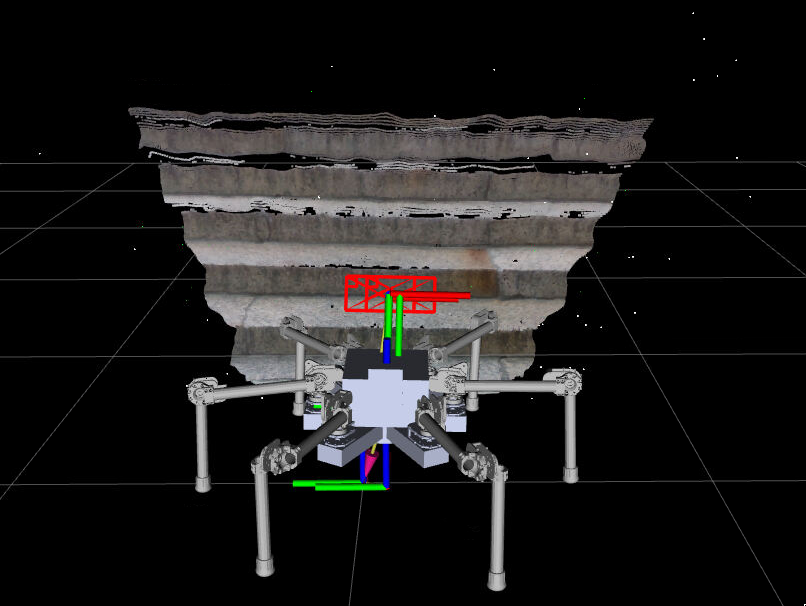}
       \caption{\emph{Left}: Hexapod robot Daisy with 18 joints actuated by series-elastic HEBI X Modules. Each module provides measurements including angular position, angular velocity, and joint torque. Additionally, an Intel Realsense D435i camera is mounted on top of the robot. \emph{Right}: A visualization of the robot's real-time field of view and estimated configuration.}
  \label{fig:sec1_overview}
\end{figure}

Our filter design is inspired by \cite{bloesch2013state}, but the sensor model of our filter is fundamentally different from their work. We designed our filter using a contact-centric approach instead of IMU-centric approach. 
We separate sensor readings into the prediction model and the measurement model of a Kalman filter based on the foot contact status of legs. The body velocity is updated using joint velocity measurements instead of accelerometer measurements. Additionally, we also use the contact status to limit when and how we incorporate IMU information into the filter. Since contact information plays a central role in the filter design, we call our algorithm contact-centric. The detailed design of the filter is explained in Section \ref{sec:filter}.

Furthermore, we also improve the performance of VINS using COCLO. In the original VINS, IMU measurements obtained between the arrival times of two camera images are integrated to serve as the initial value of the estimated camera transformation. This algorithm step is called IMU pre-integration \cite{qin2018vins}, and it is sensitive to large IMU noise. Because of this, the estimation precision of VINS is deteriorated when used on legged robots.
We replace this step by using the output of COCLO instead of IMU integration. The improvement is discussed in \ref{sec:vins}. 

To demonstrate the performance of COCLO, we conduct experiments on a hexapod robot in three different environments: flat ground, ramps, and stairs. We compare the estimation results of COCLO, the unmodified VINS, and our modified version of VINS against ground truth data. The experimental results show COCLO outperforms unmodified VINS by 25\%-40\% and the modified VINS has increased robustness. The experimental setup and analysis can be found in Section \ref{sec:experiments}.

\section{Related Work}
State estimation has been studied extensively for decades in the areas of wheeled and aerial robotics. The earliest approaches for wheeled robot odometry only used encoders to keep track of the robot's relative position displacement. This approach, however, suffered from wheel slippage, resulting in estimation drift over time. An alternative method termed Visual Odometry (VO) was developed \cite{hartley2003multiple, nister2004visual} to estimate relative position and orientation displacement from camera images, which have rich information and no wheel slippage problem. But the performance of VO algorithms had its own drawbacks due to constraints such as camera field of view, lighting conditions, and texture of the environment.

To further improve the performance of state estimation algorithms, Kalman Filtering based methods were used to fuse information together from multiple sensors to achieve higher overall accuracy. On wheeled robots, a Kalman Filter (KF) can use encoder measurements to update its process model, propagating the robot's motion through the environment. Then, it can use camera measurements from VO to correct its estimation in its measurement model. A particularly effective class of filter-based VO methods that fuses IMU and camera measurements to estimate robot pose is called Visual Inertial Odometry (VIO) \cite{li2012improving}. This method has been able to combat many of the problems in global drift that wheeled robots previously struggled with. 

Kalman filters were initially designed for linear systems with Gaussian noises, making them ineffective on many real robots. The Extended Kalman Filter (EKF) and the Unscented Kalman Filter (UKF) \cite{wan2000unscented} were developed to adapt Kalman Filtering to nonlinear systems. While the EKF proved to be computationally efficient, the UKF estimation accuracy was better on highly nonlinear systems \cite{thrun2005probabilistic}. To improve the efficiency of UKF, square root UKF (SR-UKF) is invented, which runs 20\% faster than standard UKF and 10\% faster than EKF \cite{van2001square}.

Depending on how sensor models are organized, VIO approaches are usually categorized as loosely or tightly coupled \cite{wendel2004tightly}. Loosely-coupled approaches average the estimations from different sensors to get the final result. On the other hand, tightly-coupled approaches use one integrated model for all of the sensors. 

Research on state estimation for legged robots started less than 15 years ago.  \cite{lin2005leg}\cite{lin2006sensor}\cite{skaff2010context} developed the earliest legged robot state estimation algorithms on RHex \cite{saranli2001rhex}. However, these works were highly constrained by RHex's mechanical design, and they could not be generalized to other robots or more difficult terrains. Since then, several advances in legged state estimation have emerged. \cite{reinstein2011dead} presented a data-driven legged odometry approach that generalized to different gaits. \cite{bloesch2013state2}\cite{bloesch2013state} implemented an EKF that fused joint encoder and IMU measurements to estimate the full state of a StarlETH quadruped robot \cite{hutter2012starleth}. \cite{ma2016real} also employed an EKF to do state estimation for the DARPA LS3 Robot. By carefully synchronizing sensors and isolating shocks on sensors, \cite{ma2016real} successfully lowered the position estimation error to 1\%  over a 300m flat ground trail. Most recently, \cite{wisth2019robust} combined IMU, encoder, and camera measurements using a factor-graph approach. Our work builds upon the ideas mentioned above and presents a filter design for legged robot. To the best of our knowledge, our work is the first to use contact information instead of IMU information as the key design factor for legged robot state estimation.

\section{Contact-Centric Leg Odometry} \label{sec:filter}
\begin{figure}
\centering
\includegraphics[width=\linewidth]{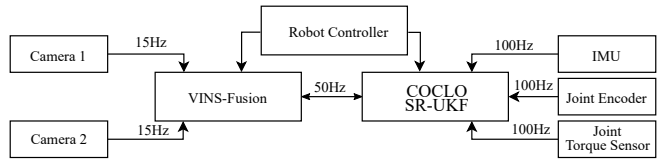}\label{fig:arch}
\caption{Block architecture of COCLO state estimation algorithm and it relationship with VINS-Fusion}
\end{figure}

COCLO uses Square Root Unscented Kalman Filter (SR-UKF) to estimate the state of the robot. In this section we first introduce necessary notations for rigid body transformations, and then we introduce the SR-UKF state, prediction model and measurement model. The filter is designed to rely mainly on joint sensor instead of IMU. 
\subsection{Notation}
Throughout the paper we will adopt the notation in \cite{murray2017mathematical} to represent vectors and rotations. A rotation matrix $R_{wc}$ changes the representation of a vector $p_c$ in CoM frame of the robot (c) into the world frame (w), denoted as
\begin{align*}
    p_w = R_{wc}p_c.
\end{align*}
We also denote the translation between world frame and CoM frame as $t_{wc}$. 

We use quaternions to parametrize rotation. The quaternion corresponds to the rotation matrix $R_{wc}$ is $q_{wc} = \begin{bmatrix} q_1 \ q_2 \ q_3 \ q_4 \end{bmatrix}$. Given a quaternion $q_{wc}$ and angular velocity $\omega_c = \begin{bmatrix} \omega_1 \ \omega_2 \ \omega_3 \end{bmatrix}^T$ in CoM frame, the update equation of the quaternion is
\begin{equation}\label{eqn:quaternion}
        \dot{q}_{wc} = \frac{1}{2}q_{wc} \otimes 
    \begin{bmatrix}
        \omega_1 \\ \omega_2 \\ \omega_3 \\ 0
    \end{bmatrix}
\end{equation}

in which $q \otimes p$ is one way to define quaternion product

\begin{align*}
    q \otimes p = 
    \begin{bmatrix}
    p_4 & p_3 & -p_2 & p_1 \\
    -p_3 & p_4 & p_1 & p_2 \\
    p_2 & -p_1 & p_4 & p_3 \\
    -p_1 & -p_2 & -p_3 & p_4 \\
    \end{bmatrix}
    \begin{bmatrix}
    q_1 \\ q_2 \\ q_3 \\q_4
    \end{bmatrix}
\end{align*}
The discretized version of Equation \ref{eqn:quaternion} is 
\begin{align}
   q_{wc(t+\Delta t)} = q_{wc(t)} + \frac{1}{2} q_{wc(t)} \otimes 
    \begin{bmatrix}
    \omega_c\Delta t \\ 0
    \end{bmatrix} 
\end{align}
For each individual leg, we model it as a fixed base manipulator and calculate the transformation matrix between the center of the foot of the leg (f) and the CoM frame of the robot using manipulator kinematics \cite{murray2017mathematical}. Given joint angles $\alpha$, which is a vector contains angles for each joint, the transformation is given by
\begin{equation}\label{eqn:fk}
  (R_{cf}, t_{cf}) = FK(\alpha)  
\end{equation}
Then position of the foot center represented in world frame is
\begin{equation*}
p_w = R_{wc}t_{cf} + t_{wc}
\end{equation*}
Further, if the joint velocity sensors on one leg give joint angular velocities $\dot{\alpha}$, then the foot velocity represented in world frame is 
\begin{equation}\label{eqn:vel}
    \dot{p}_w = R_{wc}R_{cf} J_b(\alpha) \dot{\alpha}
\end{equation}
where $J_b()$ is the body manipulator Jacobian.

\subsection{State Definition}
We use SR-UKF \cite{van2001square} to estimate the state $x_t$ and the Cholesky decomposition of its covariance $S_t$ (also referred as Cholesky factor). 

The state is described as a vector of dimension 46 for 6-legged robot: 
\begin{align*}
    x_t := \Big[ \{r_w, v_w\}, \{q_{wc}, \omega_w\}, \{b^g, b^a\}, \{p^1_w \dots  p^n_w\}, \{c^1 \dots c^n\} \Big]^T
\end{align*}
where
\begin{itemize}
    \item $\{r_w, v_w\}$ are the CoM position and velocity of the robot represented in world frame.
    \item $\{q_{wc}, \omega_w\}$ are the quaternion representing the rotation from robot frame to world frame and its angular velocity represented in world frame.
    \item $\{b^g, b^a\}$ are the estimated gyroscope and accelerometer biases.
    \item $\{p^1_w \dots p^n_w\}$ is the list of foot positions also in world frame.
    \item $\{c^1 \dots c^n\}$ is the list of foot contact status, which represents the probability that each of the foot is in contact with ground. For $i \in [1, n]$, $c^i \in [0, 1]$.
\end{itemize}

Notice that the state definition does not make assumption of number of legs so it can be easily extend to legged robots with different number of legs. 

\subsection{Prediction Model}\label{sec:prediction}
The prediction model does not take in IMU data. For body position and orientation we use constant velocity model to update their predictions. For the legs, if a leg is in stance status we do not update its position and we give it low prediction noise. While for swing legs, we use joint velocities to calculate their foot velocities. So the foot position can be updated using foot velocities. Therefore, the inputs of the prediction model at time t only contain velocities of each foot $\dot{p}^1_t \dots \dot{p}^n_t$ calculated from Equation \ref{eqn:vel}. For a time interval $\Delta t$, the prediction model predicts the updated state as follows:
\begin{align}
& r_{t+1} = r_{t} + v_{t}\Delta t \nonumber\\
& v_{t+1} = v_{t} \nonumber\\
& q_{t+1} = q_{t} + \frac{1}{2} q_{t} \otimes 
    \begin{bmatrix}
    (\omega_t - b^g_t)\Delta t \\ 0
    \end{bmatrix} \nonumber\\
& \omega_{t+1} = \omega_t, \ \ b^g_{t+1} = b^g_t, \ \  b^a_{t+1} = b^a_t \nonumber\\
& p^i_{t+1} = p^i_t +  \textbf{isSwing}(c^i_t)\dot{p}^i\Delta t, i \in [1, n] \\
& c^i_{t+1} = c^i_{t}, i \in [1, n] \nonumber
\end{align}
In Equation 5, $\textbf{isSwing}(c^i_t)$ is a boolean function that ensures foot positions are only updated in swing phase when $c^i_t$ is smaller than a threshold. This prediction equation enables precision estimation of foot positions during swing. Moreover, terms in the prediction noise covariance matrix $Q$ correspond to swing legs increase while terms related to stance legs decrease.
\subsection{Measurement Model}\label{sec:measurement}
The measurement model $h(x)$ outputs a vector contains gravity direction measurement, CoM linear velocity and angular velocity of the robot, forward kinematics of each leg as well as contact status of each foot. We represent measurement as 
\begin{align*}
h(x_t) = 
[ q^T_t g + b^a_t, \enskip v_t, \enskip \omega_t, \enskip q_t^T(r_t - p_t^i) , \enskip c_t^i \ \ \ (i=1\dots n)]^T 
\end{align*}
The measurement vector has dimension $32$.
\begin{figure}
    \centering
    \includegraphics[width=\linewidth]{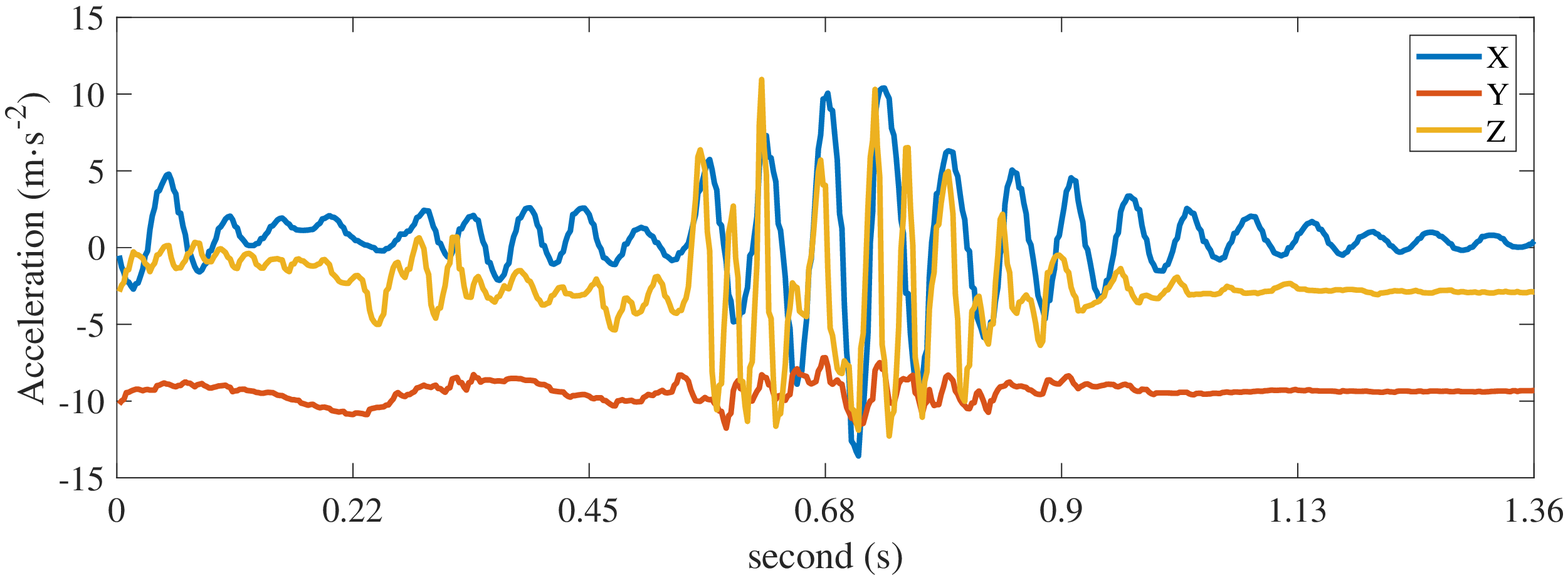}
    \caption{Accelerometer noise of Intel Realsense D435i mounted on Daisy robot during walking. Although the gravity is exerted along y axis, a significant noise along x and z is also appeared. These low-frequency but high-energy noise challenges traditional filtering and VIO methods.}
    \label{fig:acc_noise}
\end{figure}
Note that the gravity direction measurement term is commonly used in IMU complementary filters \cite{madgwick2010efficient}. But we only use the measurement when all legs have contact with the ground, other times we set the innovation to be zero. Moreover, the current accelerometer reading is averaged with previous reading to filter out high frequency noise. 

The CoM velocity and angular velocity of the robot is measured from the joint encoders and joint velocity sensors. When the robot body moves with linear velocity $v$ and angular velocity $\omega$, the velocity of one stance foot, which is stationary in world frame, relative to CoM is $-(v + \omega \times r)$ where $r$ is the distance from CoM to the tip of the foot. For a leg $i$ with joint angle $\alpha_i$ and joint angular velocity $\dot{\alpha}_i$, we can calculate $r$ using the forward kinematics (Equation \ref{eqn:fk}) of leg i that outputs relative rotation and translation between foot frame and body frame. The velocity of one foot $\dot{p}$ represented in CoM frame is calculated using Equation \ref{eqn:vel}. Given measurements from $m$ stance legs, We can write
\begin{align} \label{eqn:vw}
    \begin{bmatrix}
      I  \ \ -\lfloor r_{1\times} \rfloor \\
            \vdots \\
      I  \ \ -\lfloor r_{m\times} \rfloor
    \end{bmatrix}
    \begin{bmatrix}v \\ \omega \end{bmatrix} = 
    \begin{bmatrix}
              q(\dot{p}_1) \\
            \vdots \\
               q(\dot{p}_m) 
    \end{bmatrix}
\end{align}
in which $\lfloor r_{1\times} \rfloor$ converts vector $r_1$ to a skew symmetric matrix, and $q(p)$ uses quaternion $q$ to rotate vector $p$. We then solve Equation \ref{eqn:vw} as a least square problem to approximate $v$ and $\omega$. The solved $\omega$ is then averaged with current gyroscope reading to get the final angular velocity measurement.

The measurement for $q_t^T(r_t - p_t^i)$ is just $r$ discussed above calculated from forward kinematics $FK(\alpha)$. 

The contact status of each foot is determined using joint torque sensors that commonly available on serial-elastic actuators. We transform the torques experienced on joints, denoted as $\tau$, into foot force $F$ follows
\begin{align}
F = q(R J_b^{-T}(\alpha)\tau)
\end{align}
We record force experienced during normal walks to get $F_{max}$ and $F_{min}$ for stance legs, then we use $\frac{\|F|}{\|F_{max}\|-\|F_{min}\|}$ as the probability of contact in order  to estimate foot contact status.

\subsection{Discussion}
The key improvement we have over previous work \cite{bloesch2013state} is to use joint velocity information instead of accleerometer reading to infer body velocity. The reason why we only use accelerometer data in the measurement model when all feet are on the ground is shown in Figure \ref{fig:acc_noise}. Due to the serial elasticity in robot joints, foot impact with ground will generate large noise to corrupt accelerometer signal. So conventional IMU-centric filter will easily get affected. 

The value of prediction noise matrix Q and measurement noise matrix R of the SR-UKF are empirically tuned. Moreover, during filter updates, whenever we are certain that a leg is in contact with ground, the corresponding noise terms in Q are reduced by 2 orders of magnitude. 

\section{Modified VINS} \label{sec:vins}
Visual Inertial Navigation System (VINS), as one of start-of-the-art VIOs, is an IMU-centric method because it relies heavily on IMU pre-integration in its calculation of robot state \cite{qin2018vins}. Because IMU data on legged systems suffers from high energy noise, we propose replacing the traditional IMU pre-integration step in VINS with our COCLO output. The modifies VINS has better robustness against large IMU shock.

During a certain period of time, multiple camera images and IMU measurements are put in a factor graph optimization framework. A transformation can be solved by visual information and the same transformation can also be obtained by integrating the IMU measurements between two camera images. Therefore the two information can be jointly optimized to get a better estimation. Consider time instances $t_k$ and $t_{k+1}$ that correspond to the arrival times of two image frames, IMU measurements in between are used to update the position and velocity estimation. Assume the position and velocity estimation at $t_k$ are $p^w_{t_k}$ and $v^w_{t_k}$, then the next position estimation $p^w_{t_{k+1}}$ is \cite{qin2018vins}
\begin{align*}
    p^w_{t_{k+1}} =  p^w_{t_{k}} & + v^w_{t_k}\Delta t \\
                    & + \iint_{t\in[t_{k}, t_{k+1}]} (R_t(a-b_a) - g_w)dt^2
\end{align*}
The integration step assumes that the accelerometer measurement only comes from the motion of the VIO sensor, which is not the case for legged robots. As mentioned in Figure \ref{fig:acc_noise}, the noise will enter the integration result and make integrated position displacement deviates from the true value. 
Therefore IMU pre-integration is not accurate. The sliding window and loop closure steps \cite{qin2018vins} in VINS can mitigate this problem. But large IMU shock will make VINS perform worse in certain cases as we will show in Section \ref{sec:experiments}. 

Instead, with COCLO runs SR-UKF in parallel with the VINS, we can use the output of SR-UKF as position estimation $p^w_{t_{k+1}}$ and orientation estimation for VINS. At the same time, the output of the VINS can also be used to perform a measurement update for the position, velocity and orientation of the SR-UKF state. 


\section{Experiments}\label{sec:experiments}
\begin{figure}
\centering
\includegraphics[width=0.8\linewidth]{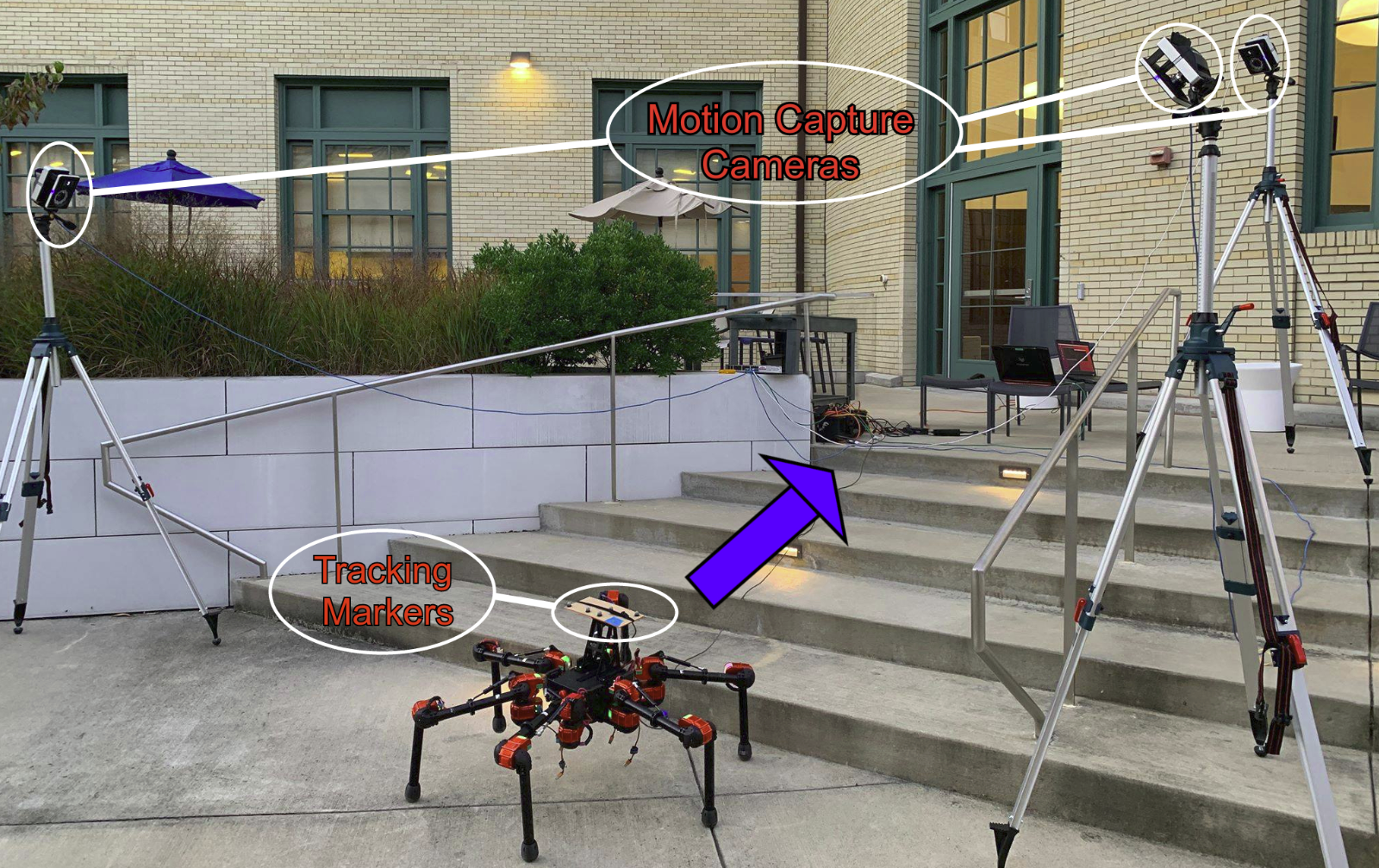}
\caption{Experimental setup when testing on stairs. Daisy is about to climb a 5-step stair where each of the steps has a width of 0.6m and a height of 0.15m (14.04\degree incline). 3 Vicon cameras are set up to record ground the truth location and orientation of the robot.}
\end{figure}
Our experimental platform is a hexapod robot, Daisy, that consists of 18 series-elastic actuators \cite{pratt1995series}. Each actuator contains an IMU, an encoder, and a torque sensor. Data from these sensors is processed at 100 Hz on-board an Intel i7 NUC mini PC. To validate the performance of COCLO, we implemented an IMU-centric filter \cite{bloesch2013state} on Daisy, but we were unable to get reasonable results due to high energy IMU noise during walking. Instead, we compare COCLO with VINS-Fusion\footnote{\url{https://github.com/HKUST-Aerial-Robotics/VINS-Fusion}}, an open source implementation of VINS. To gather visual information, we mounted an Intel Realsense D435i depth camera on top of Daisy to collect stereo images at 15 Hz and camera IMU data at 100 Hz.

While all components of our proposed state estimator can run on-board in real time, during our experiments we only collected raw data. We ran our state estimation pipeline on this data offline so that we could easily tune filter parameters and compare the performance of different algorithms.

\begin{figure}
\centering
\includegraphics[width=0.8\linewidth]{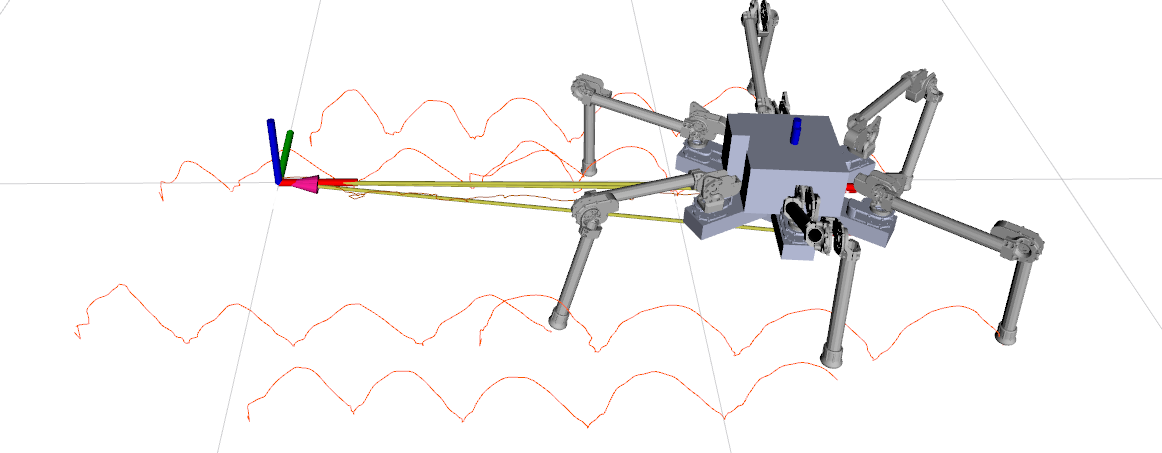}
\caption{Visualization of COCLO estimation results. The robot estimates its body pose and foot positions during walking.}
\end{figure}

\subsection{Experimental Procedures}

During our experiments, we drove the robot over flat ground, ramps, and stairs while simultaneously collecting the robot's ground truth pose and sensor measurements. The ground truth measurements were obtained using three Vicon motion capture cameras, covering the entire testing area. For the flat ground trials, our robot completed a 6m long square trajectory. We conducted the ramp trials on an outdoor ramp with 3.34m width and 0.98m elevation (16.35\degree slope). Lastly, for the stair trials, we drove the robot to climb up 5 steps, each of 0.6m width and 0.15m height (14.04\degree slope).

The robot controller we implemented on Daisy for our experiments takes high-level directional commands from an externally operated joystick. From these commands, an implementation of wave gait running on-board moves the robot in the desired direction and keeps the robot balanced. Using force feedback, this low level controller also allows the robot to feel stairs and safely navigate over them. 


\subsection{COCLO and VINS-Fusion comparison}
We first compare the performance of COCLO against unmodified VINS-Fusion and ground truth data on flat ground, ramps, and stairs. In each of our experiments in this section, COCLO and VINS-Fusion run independent of each other. 

\begin{figure}
    \centering
    \includegraphics[width=\linewidth]{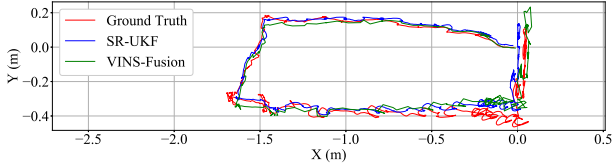}
    \caption{Flat ground X \& Y position estimation comparison.}
    \label{fig:flat_ukf_vins}
\end{figure}

During our flat ground experiment, the robot initially moved in a square trajectory with a constant speed. At the last turn, however, the robot was driven back and forth to create unstable motion and test the stability of our filter. The traveled trajectory is shown in Figure \ref{fig:flat_ukf_vins}, and it can be seen that both COCLO and VINS-Fusion estimate position trajectory reasonably well. 
After the robot traveled one meter, COCLO's estimation drift was 1.01\% and VINS-Fusion's estimation drift was 1.65\%. 

\begin{figure}
    \centering
    \includegraphics[width=\linewidth]{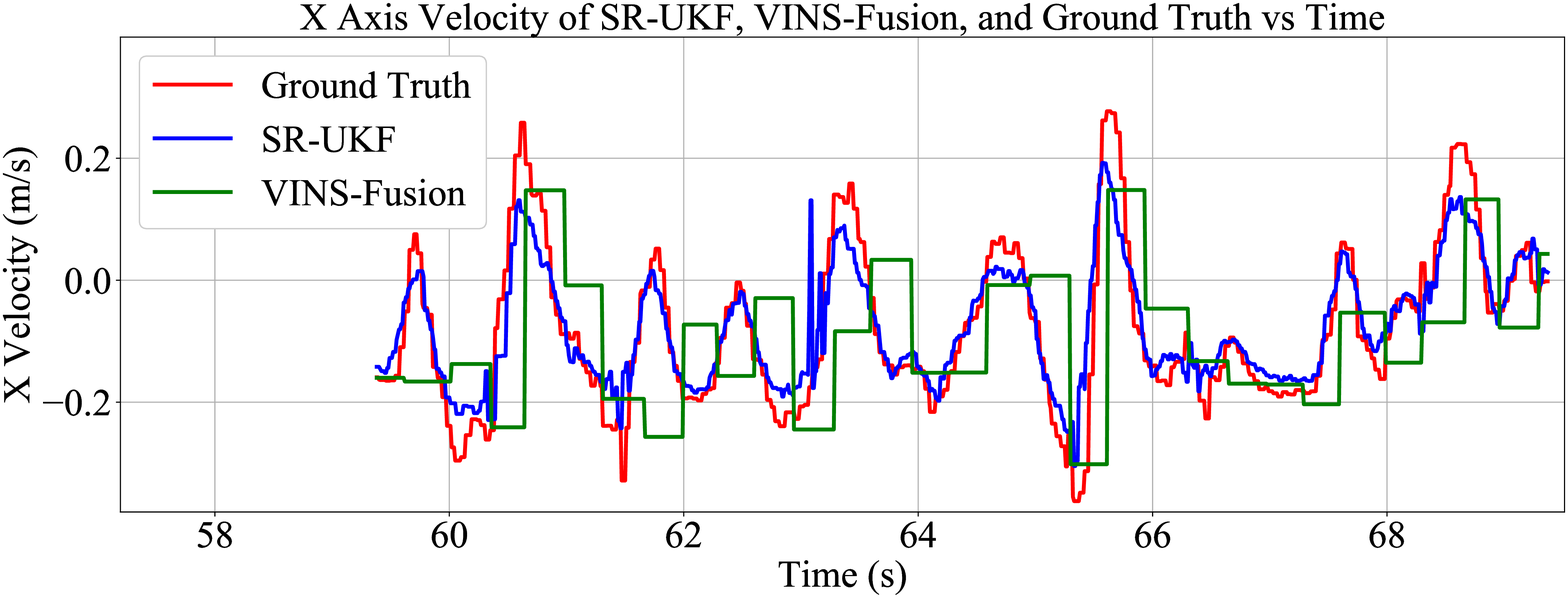}
    \includegraphics[width=\linewidth]{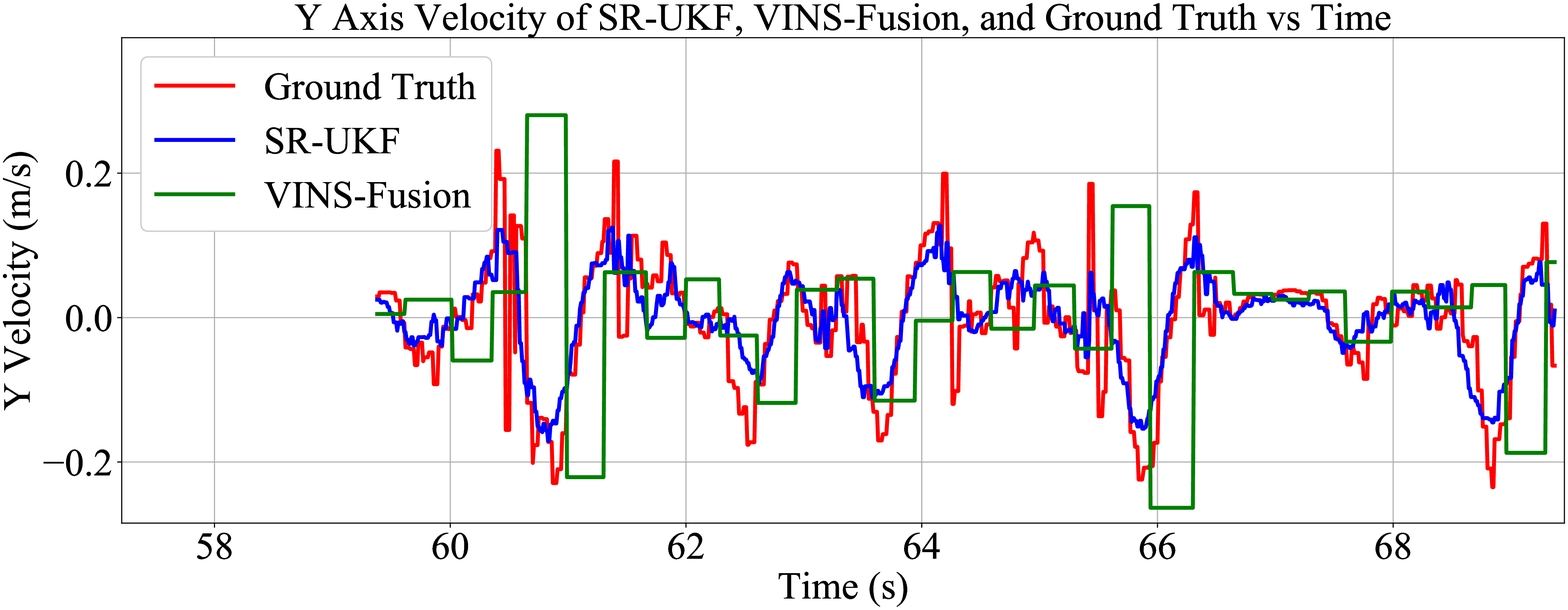}
    \caption{Flat ground estimated velocity comparison. The VINS-Fusion runs only at 2-3Hz so the VINS-Fusion velocity between two updates stays the same.}
    \label{fig:velocity_compare}
\end{figure}

Moreover, we also compare the velocity estimation of COCLO and VINS-Fusion. Figure \ref{fig:velocity_compare} compares the estimated velocity along the X and Y axes. It can be seen that COCLO estimates velocities with higher precision than VINS-Fusion. The result also partially explains why the COCLO has smaller position drift than VINS-Fusion.

\begin{figure}
    \centering
    \includegraphics[width=\linewidth]{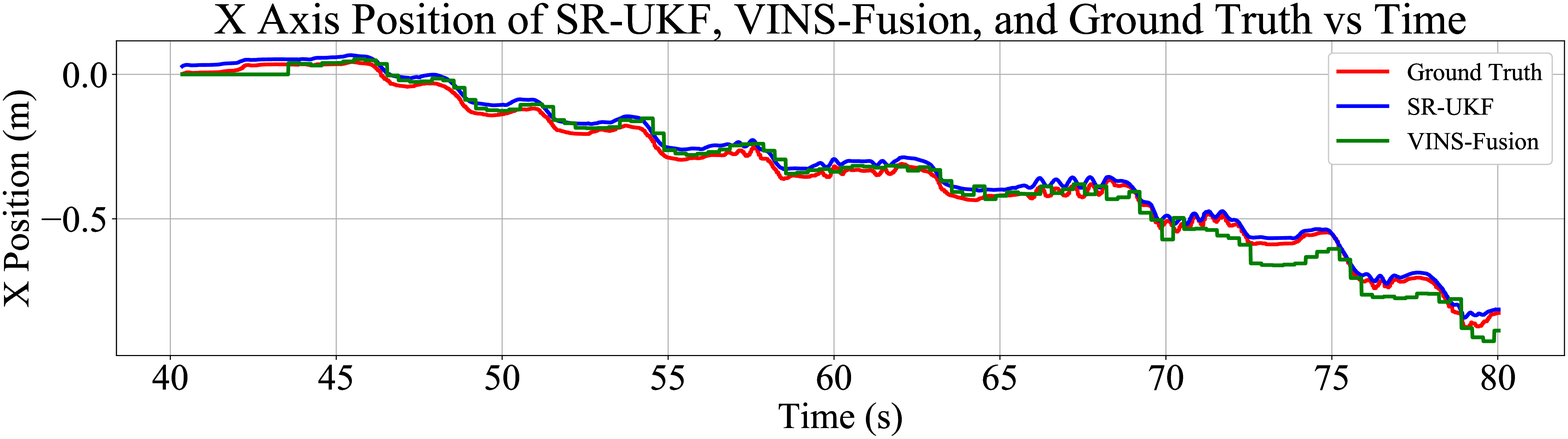}
    \includegraphics[width=\linewidth]{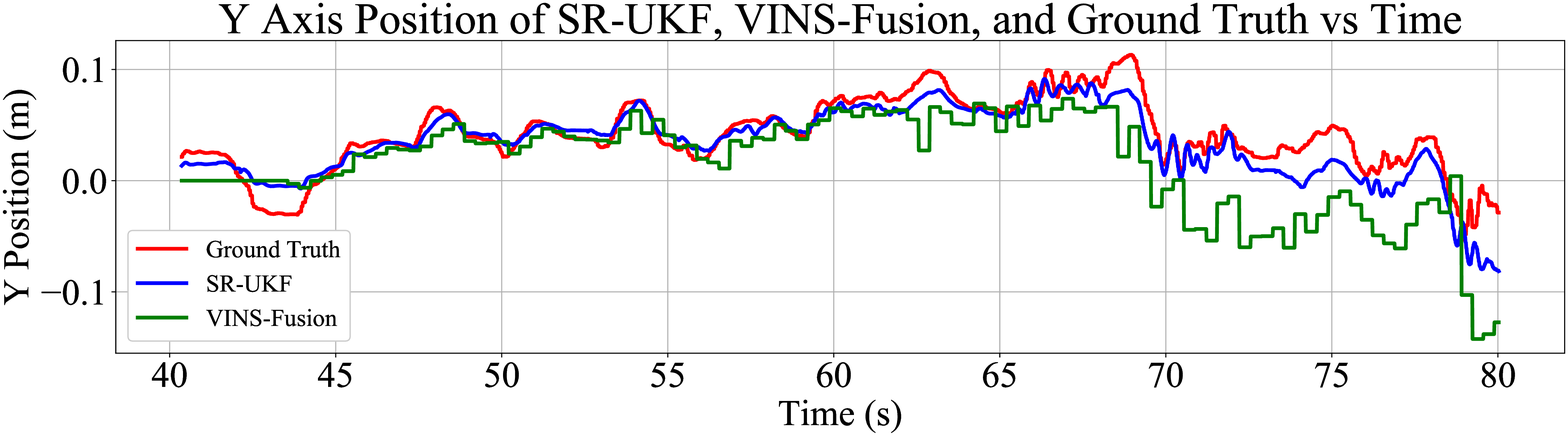}
    \includegraphics[width=\linewidth]{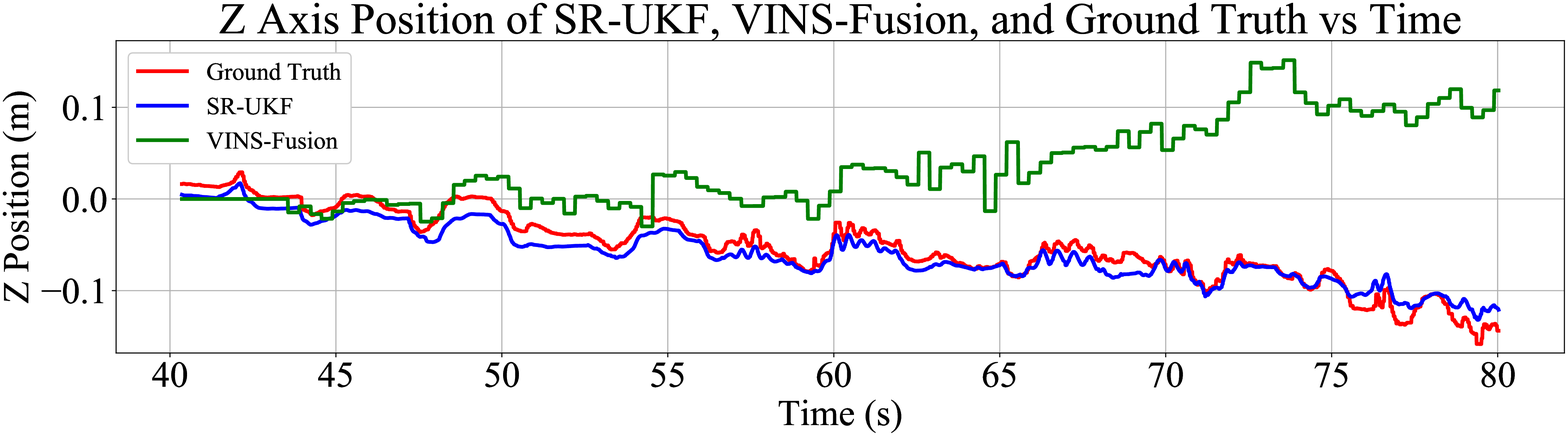}
    \caption{Algorithm runs on data recorded during a ramp trial. (Top) Estimated X position (Middle) Estimated Y position (Bottom) Estimated Z position. The severe shaking and feature less ramp surface makes VINS-Fusion Z direction estimation drift away.}
    \label{fig:ramp_pos_compare_combine}
\end{figure}

For our second experiment we compare the performance of COCLO and VINS-Fusion on a ramp. In one ramp trial, we let the robot climb up the ramp and manually shake the robot as external disturbance. Figure \ref{fig:ramp_pos_compare_combine} presents the estimated position in all directions.

\begin{figure}
    \centering
    \includegraphics[width=\linewidth]{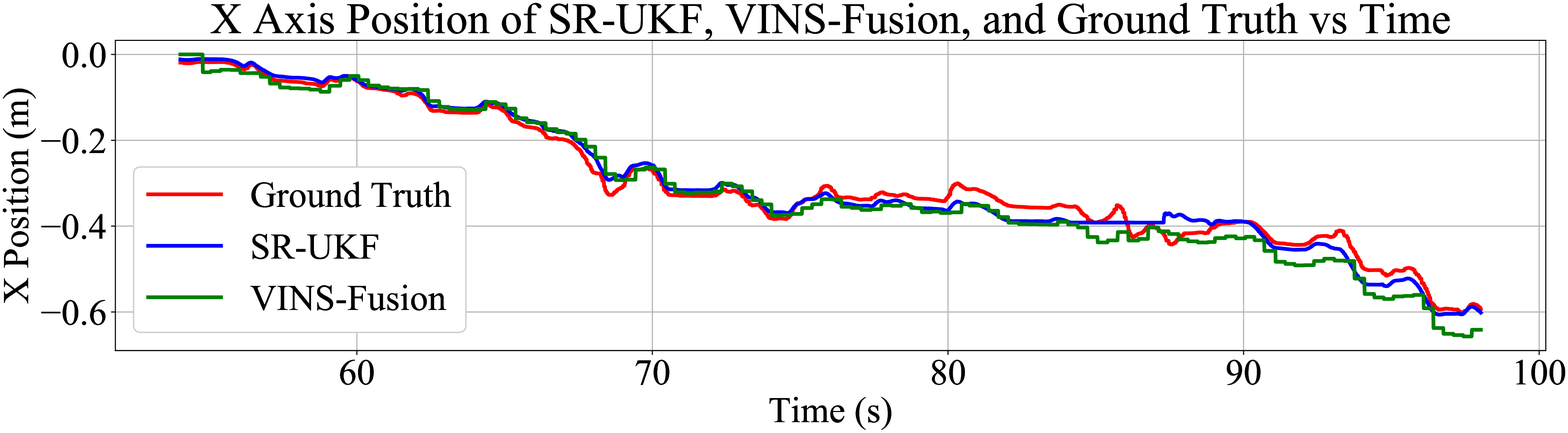}
    \includegraphics[width=\linewidth]{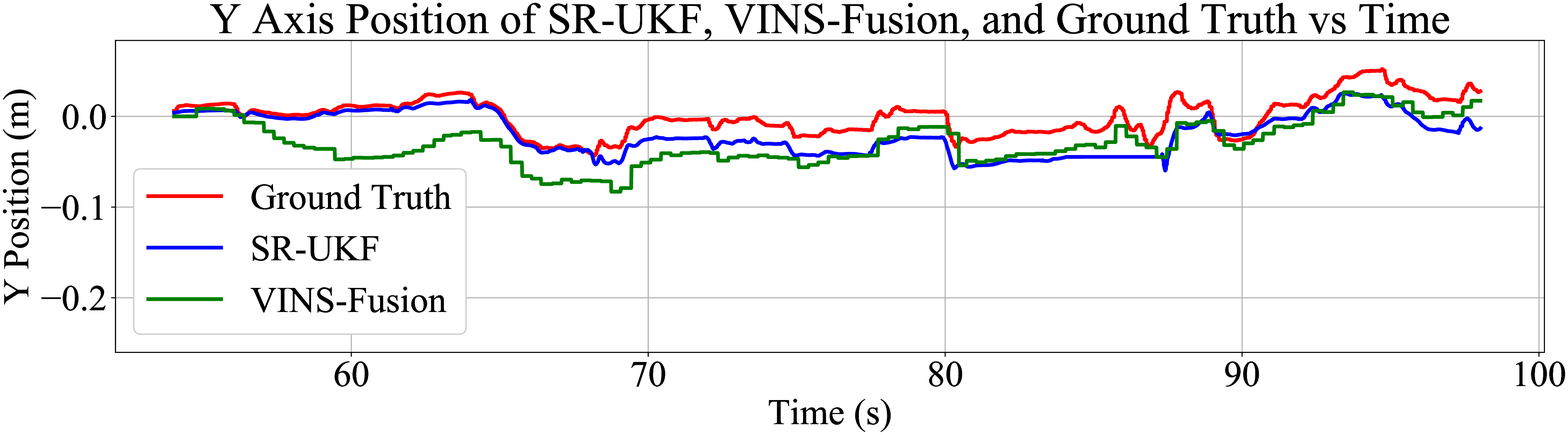}
    \includegraphics[width=\linewidth]{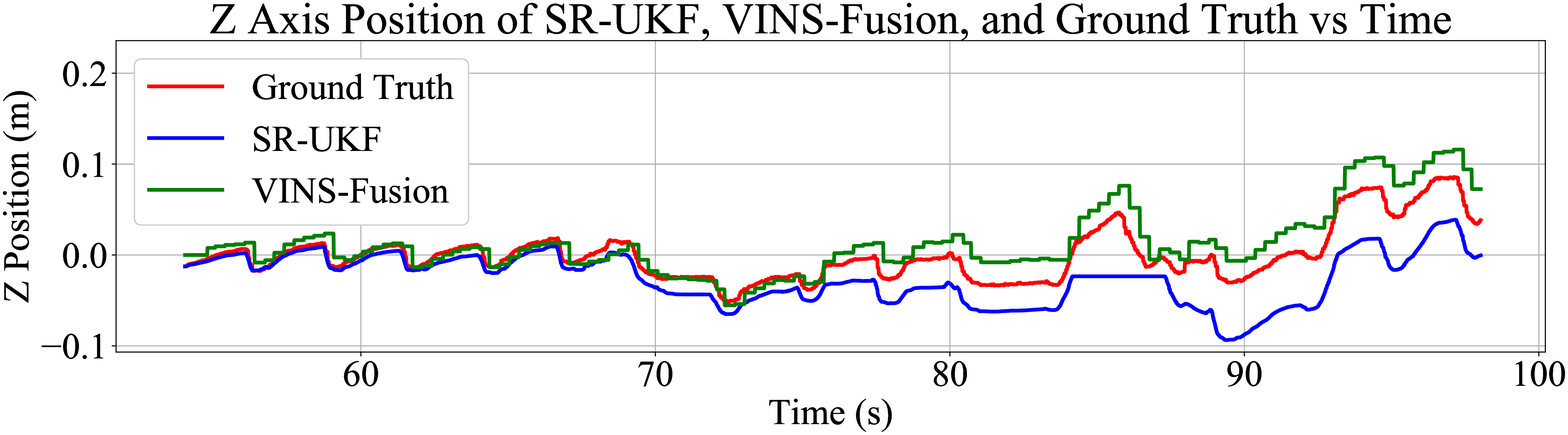}
    \caption{Algorithm runs on data recorded during a stair trial in which \textbf{COCLO and VINS-Fusion are separate}. (Top) Estimated X position (Middle) Estimated Y position (Bottom) Estimated Z position}
    \label{fig:stair_pos_compare}
\end{figure}

In the stair trials, the robot has to deal with more challenging conditions. Figure \ref{fig:stair_pos_compare} presents the estimated position in all directions during one trial. In the X direction, both COCLO and VINS-Fusion stay close to the ground truth measurement over the entire 0.6 meter travel distance. However, since the robot's motion is very unstable when climbing up stairs, both algorithms have large Y direction drifts. In the Z direction, VINS-Fusion has smaller drift than COCLO. Even though COCLO only uses accelerometer measurements to update its velocity prediction when all legs are on the ground, shock with large acceleration in Z direction still exists in transient period. In VINS-Fusion, this drift may come from both IMU shock and poor feature tracking when the camera faces regions with less features on stairs. 

\subsection{COCLO + VINS-Fusion Experiment}
\begin{figure}
    \centering
    \includegraphics[width=\linewidth]{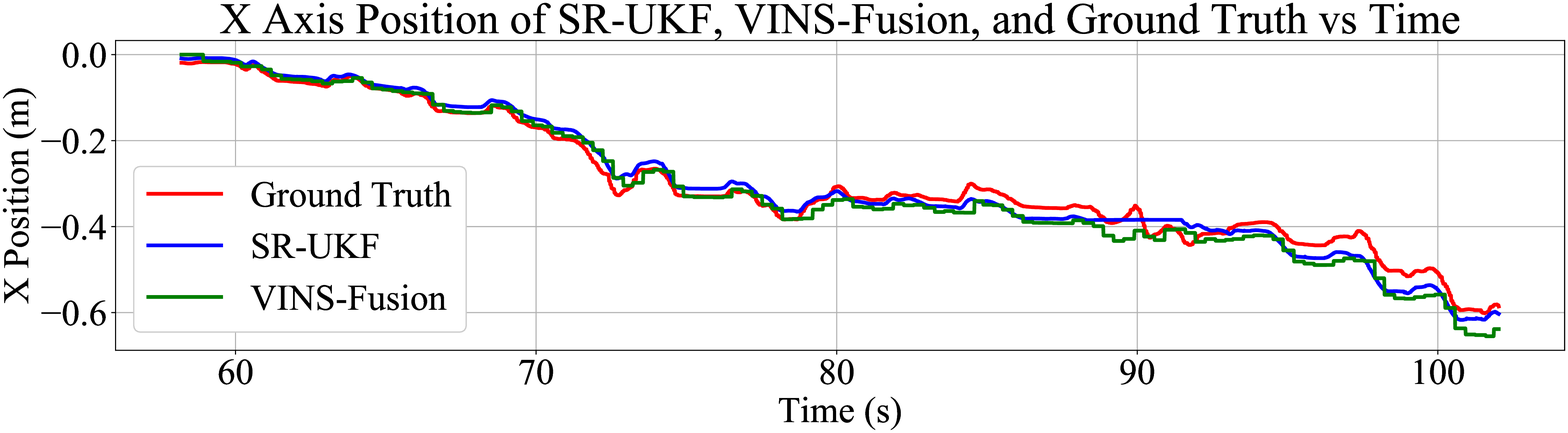}
    \includegraphics[width=\linewidth]{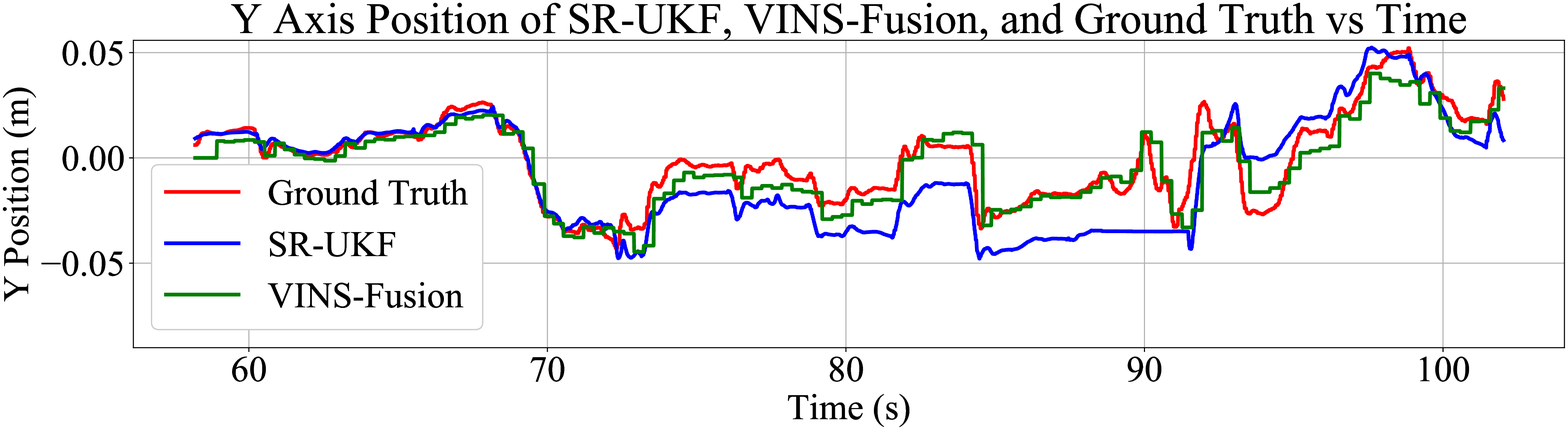}
    \includegraphics[width=\linewidth]{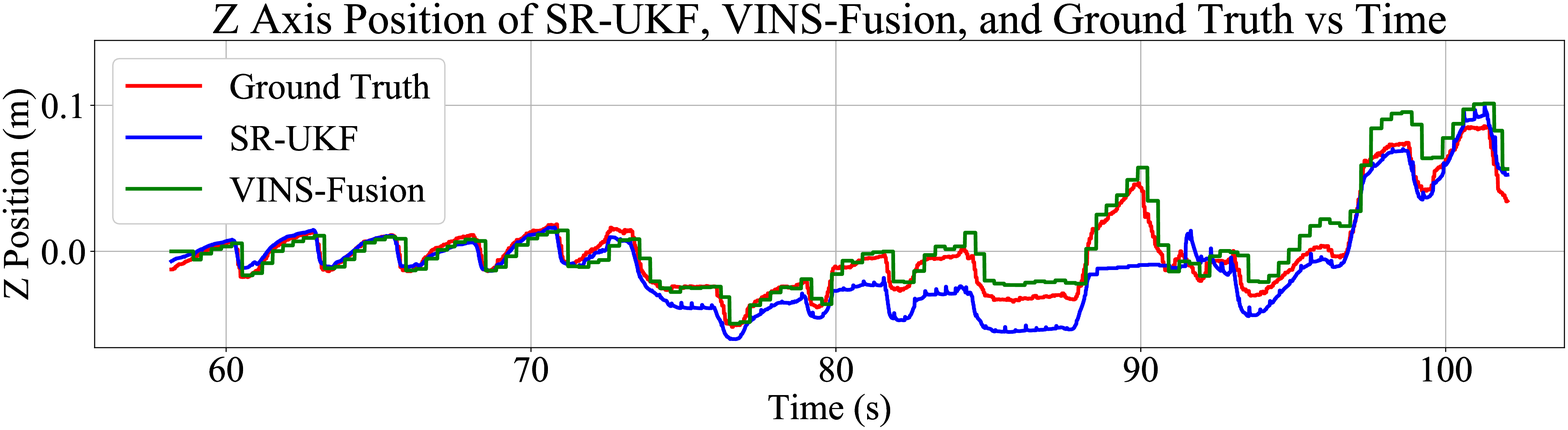}
    \caption{Algorithm runs on data recorded during the same stair trial in which \textbf{COCLO and VINS-Fusion are combined}. (Top) Estimated X position (Middle) Estimated Y position (Bottom) Estimated Z position}
    \label{fig:stair_pos_compare_combine}
\end{figure}

By applying a modification to VINS-Fusion discussed in Section \ref{sec:vins}, we can achieve better state estimation on flat ground and stairs. We implementd the combined state estimator and ran the algorithms again on the dataset recorded in same trials. Figure \ref{fig:stair_pos_compare_combine} shows the result of the same stairs climbing trial reported in the previous section. Comparing with Figure \ref{fig:stair_pos_compare}, the estimation results are improved in all three directions.

Table \ref{tab:my_label} summarizes the final drift percentage (Difference between estimated position and ground truth position divided by ground truth position) in each trials. In all test scenarios, COLCO has better performance than VINS-Fusion.

\begin{table}
    \begin{center}
     \begin{tabular}{|c | c | c|} 
     \hline
     \multicolumn{3}{|c|}{\textbf{Drift Comparison for Different Trials}} \\ 
     \hline
     \multirow{2}{*}{Flat} & COCLO & \textbf{1.01\%} \\ 
     & VINS-Fusion & 1.65\% \\ 
     \hline
     \multirow{2}{*}{Flat} & Combined COCLO & 1.57\% \\ 
     & Combined VINS-Fusion & \textbf{1.46}\% \\ 
     \hline
     \multirow{2}{*}{Ramp} & COCLO & \textbf{6.51\%} \\ 
     & VINS-Fusion & 22.04\% \\ 
     \hline
     \multirow{2}{*}{Stair} & COCLO & \textbf{7.28\%} \\   
     & VINS-Fusion & 8.55\% \\  
     \hline
     \multirow{2}{*}{Stair} & Combined COCLO & \textbf{2.61\%} \\ 
     & Combined VINS-Fusion & 5.28\% \\ 
     \hline
     \end{tabular}
    \end{center}
    \caption{Summarization of estimator drifts for different trials. }
    \label{tab:my_label}
\end{table}

\section{Conclusions}
In this paper, we introduced a state estimation algorithm called COCLO. COCLO fuses proprioceptive sensors that are commonly available on legged robots. Specifically, we leveraged joint torque sensors in serial-elastic actuators to design a contact-centric SR-UKF. This contact-centric approach estimate velocity more precisely even with the presence of jittery motion and large IMU shock. Thus COCLO enables accurate state estimation on different terrains. Furthermore, COCLO can be used to improve the performance of the state-of-the-art VINS estimator. Through experiments, we verified our filter design on flat ground, ramp, and stairs using a hexapod robot. In most cases, COCLO outperforms VINS-Fusion. Our work shows that for legged robots, contact-centric filter may be a better choice than IMU-centric approaches because joint sensors are less prone to foot impact and external disturbances. For future works, we will build better foot force model and use foot contact sensors to improve foot contact status estimation, to help robot deal with foot slippage. Also we will incorporate COCLO into VINS or other visual state estimator using a tightly-coupled approach.


\section*{ACKNOWLEDGMENT}
The authors would like to thank Hebi Robotics team members for their assistance on building and maintaining the Daisy robot. The authors also appreciate Anton Egorov's help while conducting the experiments.


\bibliographystyle{plain}
\bibliography{sections/references}

\begin{thebibliography}{10}

\bibitem{bloesch2013state2}
Michael Bloesch, Christian Gehring, P{\'e}ter Fankhauser, Marco Hutter, Mark~A
  Hoepflinger, and Roland Siegwart.
\newblock State estimation for legged robots on unstable and slippery terrain.
\newblock In {\em International Conference on Intelligent Robots and Systems},
  pages 6058--6064, 2013.

\bibitem{bloesch2013state}
Michael Bloesch, Marco Hutter, Mark~A Hoepflinger, Stefan Leutenegger,
  Christian Gehring, C~David Remy, and Roland Siegwart.
\newblock State estimation for legged robots-consistent fusion of leg
  kinematics and imu.
\newblock {\em Robotics}, 17:17--24, 2013.

\bibitem{hartley2003multiple}
Richard Hartley and Andrew Zisserman.
\newblock {\em Multiple View Geometry in Computer Vision}.
\newblock Cambridge University Press, 2003.

\bibitem{hutter2012starleth}
Marco Hutter, Christian Gehring, Michael Bloesch, Mark~A Hoepflinger, C~David
  Remy, and Roland Siegwart.
\newblock Starleth: A compliant quadrupedal robot for fast, efficient, and
  versatile locomotion.
\newblock In {\em Adaptive Mobile Robotics}, pages 483--490. 2012.

\bibitem{li2012improving}
Mingyang Li and Anastasios~I Mourikis.
\newblock Improving the accuracy of ekf-based visual-inertial odometry.
\newblock In {\em International Conference on Robotics and Automation}, pages
  828--835, 2012.

\bibitem{lin2005leg}
Pei-Chun Lin, Haldun Komsuoglu, and Daniel~E Koditschek.
\newblock A leg configuration measurement system for full-body pose estimates
  in a hexapod robot.
\newblock {\em IEEE Transactions on Robotics}, 21(3):411--422, 2005.

\bibitem{lin2006sensor}
Pei-Chun Lin, Haldun Komsuoglu, and Daniel~E Koditschek.
\newblock Sensor data fusion for body state estimation in a hexapod robot with
  dynamical gaits.
\newblock {\em IEEE Transactions on Robotics}, 22(5):932--943, 2006.

\bibitem{ma2016real}
Jeremy Ma, Max Bajracharya, Sara Susca, Larry Matthies, and Matt Malchano.
\newblock Real-time pose estimation of a dynamic quadruped in gps-denied
  environments for 24-hour operation.
\newblock {\em The International Journal of Robotics Research}, 35(6):631--653,
  2016.

\bibitem{madgwick2010efficient}
Sebastian Madgwick.
\newblock An efficient orientation filter for inertial and inertial/magnetic
  sensor arrays.
\newblock {\em Report x-io and University of Bristol}, 25:113--118, 2010.

\bibitem{murray2017mathematical}
Richard~M Murray.
\newblock {\em A Mathematical Introduction to Robotic Manipulation}.
\newblock CRC Press, 2017.

\bibitem{nister2004visual}
David Nist{\'e}r, Oleg Naroditsky, and James Bergen.
\newblock Visual odometry.
\newblock In {\em Conference on Computer Vision and Pattern Recognition}, 2004.

\bibitem{pratt1995series}
Gill~A Pratt and Matthew~M Williamson.
\newblock Series elastic actuators.
\newblock In {\em International Conference on Intelligent Robots and Systems},
  volume~1, pages 399--406, 1995.

\bibitem{qin2018vins}
Tong Qin, Peiliang Li, and Shaojie Shen.
\newblock Vins-mono: A robust and versatile monocular visual-inertial state
  estimator.
\newblock {\em IEEE Transactions on Robotics}, 34(4):1004--1020, 2018.

\bibitem{reinstein2011dead}
Michal Reinstein and Matej Hoffmann.
\newblock Dead reckoning in a dynamic quadruped robot: Inertial navigation
  system aided by a legged odometer.
\newblock In {\em International Conference on Robotics and Automation}, pages
  617--624, 2011.

\bibitem{saranli2001rhex}
Uluc Saranli, Martin Buehler, and Daniel~E Koditschek.
\newblock Rhex: A simple and highly mobile hexapod robot.
\newblock {\em The International Journal of Robotics Research}, 20(7):616--631,
  2001.

\bibitem{skaff2010context}
Sarjoun Skaff, Alfred~A Rizzi, Howie Choset, and Matthew Tesch.
\newblock Context identification for efficient multiple-model state estimation
  of systems with cyclical intermittent dynamics.
\newblock {\em IEEE Transactions on Robotics}, 27(1):14--28, 2010.

\bibitem{thrun2005probabilistic}
Sebastian Thrun, Wolfram Burgard, and Dieter Fox.
\newblock {\em Probabilistic Robotics}.
\newblock MIT press, 2005.

\bibitem{van2001square}
Rudolph Van Der~Merwe and Eric~A Wan.
\newblock The square-root unscented kalman filter for state and
  parameter-estimation.
\newblock In {\em International Conference on Acoustics, Speech, and Signal
  Processing}, volume~6, pages 3461--3464, 2001.

\bibitem{wan2000unscented}
Eric~A Wan and Rudolph Van Der~Merwe.
\newblock The unscented kalman filter for nonlinear estimation.
\newblock In {\em Adaptive Systems for Signal Processing, Communications, and
  Control Symposium}, pages 153--158, 2000.

\bibitem{wendel2004tightly}
Jan Wendel and Gert~F Trommer.
\newblock Tightly coupled gps/ins integration for missile applications.
\newblock {\em Aerospace Science and Technology}, 8(7):627--634, 2004.

\bibitem{wisth2019robust}
David Wisth, Marco Camurri, and Maurice Fallon.
\newblock Robust legged robot state estimation using factor graph optimization.
\newblock {\em IEEE Robotics and Automation Letters}, 2019.

\end{thebibliography}

\end{document}